\documentclass{emulateapj}

\begin{document}

\shortauthors{Luhman et al.}
\shorttitle{Brown Dwarf Disks in $\sigma$ Ori}

\title{Disks around Brown Dwarfs in the $\sigma$ Orionis Cluster\altaffilmark{1}}

\author{
K. L. Luhman\altaffilmark{2},
J. Hern\'andez\altaffilmark{3,4},
J. J. Downes\altaffilmark{4,5},
L. Hartmann\altaffilmark{3},
and C. Brice\~no\altaffilmark{4}}

\altaffiltext{1}
{Based on observations performed with the {\it Spitzer Space Telescope}.}

\altaffiltext{2}{Department of Astronomy and Astrophysics, The Pennsylvania
State University, University Park, PA 16802; kluhman@astro.psu.edu.}

\altaffiltext{3}{Department of Astronomy, The University of Michigan,
Ann Arbor, MI 48109.}

\altaffiltext{4}{Centro de Investigaciones de Astronom{\'\i}a, 
Apartado Postal. 264, M\'erida 5101-A, Venezuela.}

\altaffiltext{5}{Escuela de F{\'\i}sica, Universidad Central de Venezuela,
Apartado Postal 47586, Caracas 1041-A, Venezuela}

\begin{abstract}

We have performed a census of circumstellar disks around brown dwarfs in
the $\sigma$~Ori cluster using all available images from
the Infrared Array Camera onboard the {\it Spitzer Space Telescope}.
To search for new low-mass cluster members with disks, we have measured 
photometry for all sources in the {\it Spitzer} images and have identified 
the ones that have red colors that are indicative of disks.
We present five promising candidates, which may consist of two brown dwarfs, 
two stars with edge-on disks, and a low-mass protostar if they are bona fide
members. Spectroscopy is needed to verify the nature of these sources.
We have also used the {\it Spitzer} data to determine which of the 
previously known probable members of $\sigma$~Ori are likely to have disks.
By doing so, we measure disk fractions of $\sim40$\% and $\sim60$\% for 
low-mass stars and brown dwarfs, respectively. These results are similar 
to previous estimates of disk fractions in IC~348 and Chamaeleon~I, which 
have roughly the same median ages as $\sigma$~Ori ($\tau\sim3$~Myr).
Finally, we note that our photometric measurements 
and the sources that we identify as having disks differ significantly from 
those of other recent studies that analyzed the same {\it Spitzer} images. 
For instance, previous work has suggested that the T dwarf S~Ori~70 is
redder than typical field dwarfs, which has been cited as possible evidence of 
youth and cluster membership. However, we find that this object is 
only slightly redder than the reddest field dwarfs in $[3.6]-[4.5]$ 
($1.56\pm0.07$ vs.\ 0.93--1.46). We measure a larger excess in
$[3.6]-[5.8]$ ($1.75\pm0.21$ vs.\ 0.87--1.19), but the flux at
5.8~\micron\ may be overestimated because of the low signal-to-noise ratio of
the detection. Thus, the {\it Spitzer} data do not offer strong evidence 
of youth and membership for this object, which is the faintest and coolest 
candidate member of $\sigma$~Ori that has been identified to date.

\end{abstract}


\section{Introduction}
\label{sec:intro}

The lifetime of an accretion disk around a young star represents a fundamental 
constraint on the amount of time available for the formation of giant planets.
The typical lifetimes of disks are estimated by comparing the 
prevalence of disks among clusters that span a range of ages 
($\tau\sim1$--10~Myr).
By measuring disk fractions as a function of stellar mass and star-forming
environment as well as age, the influence of these two factors on disk 
lifetimes can be characterized. 

Disk fractions are usually measured through infrared (IR) photometry of
young clusters ($\lambda>3$~\micron) and the identification of the stars that
exhibit excess emission from cool dust.
Observations of this kind were first performed with the 
{\it Infrared Astronomical Satellite} and ground-based near-IR cameras 
\citep{kh95,hai01} and have made rapid progress in recent years with the
{\it Spitzer Space Telescope} \citep{wer04}. 
Because {\it Spitzer} is exceptionally well-suited for detecting 
disks around members of young clusters \citep{all04,gut04,meg04,muz04},
it has been used extensively for measuring the disk fractions of stars 
\citep{meg05,lada06,car06,sic06,her07,her07b,her08,mue07,dahm07,bar07,dam07,luh08cha}.
{\it Spitzer}'s unique sensitivity to faint IR sources has enabled 
measurements of disk fractions well into the substellar regime in
nearby star-forming regions 
\citep[$d=150$--300~pc,][]{luh05frac,luh06tau2,luh08cha,gui07}.

The cluster of young stars associated with the O star $\sigma$~Ori~A 
has been thoroughly searched for low-mass stars and brown dwarfs 
\citep[e.g.,][]{bej01,bar02,mar01,zap02a,zap02b}. As a result, it is an
attractive target for measuring the disk fraction among low-mass objects.
Several recent studies have sought to do this with {\it Spitzer}. 
\citet{her07} performed {\it Spitzer} imaging of most of the $\sigma$~Ori 
cluster and measured the disk fraction as a function of mass down to 
$\sim0.1$~$M_\odot$ for a sample of $\sim300$ probable members (see also
\citet{cab06}). 
\citet{cab07} identified a sample of brown dwarf candidates from optical and 
near-IR data and used the {\it Spitzer} images from \citet{her07} to estimate 
a disk fraction for those sources, arriving at a value of $\sim50$\%. 
Through further analysis of those {\it Spitzer} data, 
\citet{zap07} found that 6 of 12 brown dwarf candidates
exhibited excess emission at 8~\micron.
\citet{sch08} obtained deeper {\it Spitzer} images of 18 of the faintest
candidate members and reported a disk fraction of 29\%. 
In addition to measuring disk fractions, the {\it Spitzer} data have been 
used to assess the youth and membership of the coolest candidate member of 
the cluster, the T dwarf S~Ori~70 \citep{zap02a}.
\citet{zap08} found that it is redder than typical field dwarfs
in $[3.6]-[4.5]$ in the images from \citet{her07}.
Similarly, \citet{sch08} suggested that S~Ori~70 could be anomalously 
red in $[3.6]-[4.5]$ and $[3.6]-[5.8]$ based on their deeper images, 
which they attributed to a disk 
or low surface gravity. In either case, the apparent color excesses 
would comprise evidence of youth for this object, whose membership in the 
$\sigma$~Ori cluster has been questioned \citep{mar03,bur04}.

The substellar population of the $\sigma$~Ori cluster extends down to 
and below the detection limits of the images that have been obtained by 
{\it Spitzer}. 
Thus, determining whether these objects have disks is a challenging task.
For instance, \citet{zap07} and \citet{sch08}
disagree on whether the {\it Spitzer} data show evidence of disks 
for half of the sources considered by both studies. 
In this paper, we seek to accurately characterize 
the disk population among low-mass stars and brown dwarfs in $\sigma$~Ori
through an analysis of all {\it Spitzer} images of this cluster that
includes careful treatment of errors and biases. 
We begin by summarizing the {\it Spitzer} observations and
our data reduction methods
(\S~\ref{sec:obs}). We then use these data to search for new low-mass 
members of $\sigma$~Ori that have disks (\S~\ref{sec:select}) and 
to estimate the disk fraction for the substellar members of the cluster
(\S~\ref{sec:pop}).

\section{Observations}
\label{sec:obs}

\begin{figure}
\plotone{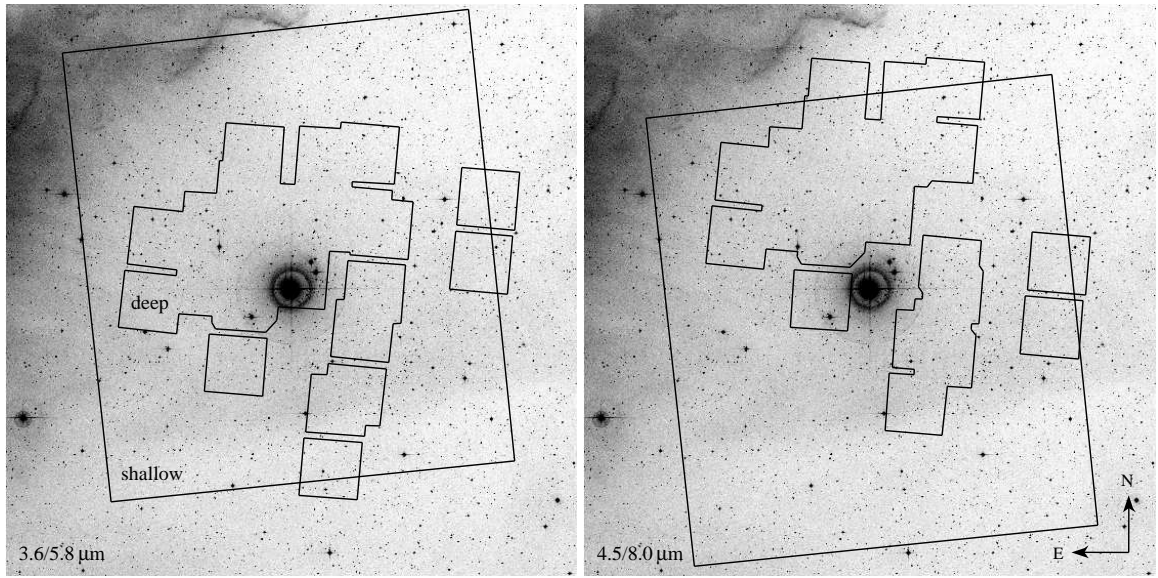}
\caption{
Fields in the $\sigma$ Ori cluster that have been imaged with IRAC.
In each of the four IRAC filters, the total exposure times were
80.4~s in the large fields \citep[''shallow",][]{her07} 
and $\geq1120$~s in the smaller regions \citep[''deep",][]{sch08}. 
The image is from DSS ($1\arcdeg\times1\arcdeg$). 
}
\label{fig:map}
\end{figure}

For our study of disks around brown dwarfs in the $\sigma$~Ori cluster, 
we use images at 3.6, 4.5, 5.8, and 8.0~\micron\ obtained with
{\it Spitzer}'s Infrared Array Camera \citep[IRAC;][]{faz04}.
We consider all IRAC observations that have been performed in this region,
which consist of images analyzed by \citet{her07} and \citet{sch08}. 
Henceforth in this paper, we refer to these sets of data as the ``shallow"
and ``deep" images, respectively. The boundaries of these images are 
indicated on images of $\sigma$~Ori from the Digitized Sky Survey (DSS)
in Figure~\ref{fig:map}.

The shallow IRAC images were collected on 2004 October 9 as a part of 
G. Fazio's IRAC Guaranteed Time Observations in {\it Spitzer} 
program 37 \citep{her07}.
Each IRAC mosaic for a given filter covered an area of 0.56~deg$^2$ 
($0.7\arcdeg\times0.8\arcdeg$). The overlapping area between the four
filters was 0.5~deg$^2$ ($0.7\arcdeg\times0.7\arcdeg$).
At each cell in the mosaic and in each filter, IRAC obtained three 1~s
exposures and three 26.8~s exposures. 

The deep IRAC imaging was executed on 2006 September 28, 2007 March 30-31, 
and 2007 April 3-4 through {\it Spitzer} program 30395 \citep{sch08}.
Instead of mosaics, 13 individual pointings were selected to encompass 18
candidate substellar members of $\sigma$~Ori. Some of the images from these
separate observations overlapped with each other, as shown in 
Figure~\ref{fig:map}. For each filter, a total area of 0.20~deg$^2$ was imaged.
An area of 0.12~deg$^2$ was covered by all four bands.
At each of the 13 pointings and in each of the 3.6, 4.5, and 
5.8~\micron\ filters, 12 images were obtained with exposure times of 96.8~s. 
For 8.0~\micron, the number of images was doubled and the exposure times 
were 46.8~s. 

The shallow images were processed with the Spitzer Science Center (SSC)
S14.0.0 pipeline and the deep data were processed with the S14.4.0 and S15.3.0 
pipelines. We combined the images produced by the SSC pipelines into mosaics and
measured photometry for all point sources appearing in them using the
methods described by \citet{luh08cha}. 
We selected a plate scale of $0.86\arcsec$~pixel$^{-1}$ for the reduced IRAC
mosaics, which is the native scale divided by $\sqrt{2}$.
We used an aperture radius of 4 pixels when measuring photometry for most
sources. Smaller apertures with radii of 2 or 3 pixels were applied to 
sources that were near other stars, which included the candidate brown dwarfs
S~Ori~25, S~Ori~45, S~Ori~47, S~Ori~54, S~Ori~65, S~Ori~68, 
S~Ori~J053932.4$-$025220, S~Ori~J053949.5$-$023130, and
S~Ori~J053929.4$-$024636. 
We also used a small aperture of 2 pixels for the new candidate member
IRAC J05384729$-$0235194 (see \S~\ref{sec:cand}).
The inner radius of the sky annulus was the same as the aperture radius in
all cases. For the 4 pixel apertures at 5.8 and 8.0~\micron, 
we selected relatively large widths of 6 pixels for the sky annuli 
to better measure the bright background emission at those wavelengths. 
In all other cases, the width of the annulus was 1 pixel.
As noted above, some of the deep images overlapped with each other. 
For stars that were observed more than once during the spring 
observations, which spanned only a few days, we have adopted the average of
the multiple measurements. Data obtained in both the fall and spring for a 
given star are presented separately. 
The completeness limits of the shallow images are 17.25, 17, 14.75, and 14 
at 3.6, 4.5, 5.8, and 8.0~\micron, respectively.
The limits for the deep images are fainter by 0.75--1.25~mag.
The quoted photometric errors include the Poisson errors in the source and 
background emission and the 2\% uncertainty in the calibration of IRAC 
\citep{rea05}.

\section{Searching for New Members with Disks}
\label{sec:select}

\subsection{Compilation of Probable Members}

In the first stage of our analysis of the disk population in $\sigma$~Ori,
we use our catalog of all sources detected by IRAC to search for new 
disk-bearing members of the cluster.
We can use the IRAC data for known probable members of $\sigma$~Ori
to illustrate the typical colors of young stars with disks.
To create a list of probable members, we begin with the sources that
were considered by the other recent studies of disks in $\sigma$~Ori 
\citep{her07,cab07,zap07,sch08}. 
We exclude sources from \citet{her07} that are resolved as galaxies in 
images from the United Kingdom Infrared Telescope (UKIRT) 
Infrared Deep Sky Survey \citep[UKIDSS,][]{law07}\footnote{
The UKIDSS project is described by \citet{law07}. UKIDSS uses the
UKIRT Wide Field Camera \citep[WFCAM,][]{cas07} and a photometric
system described by \citet{hew06}. The pipeline processing and
science archive are described by Irwin et al. (in preparation) and
\citet{ham08}. We have used data from the first data release, which is
described in detail by \citet{war07}.} 
or that have been spectroscopically classified
as nonmembers by \citet{ken05} and \citet{sac08}.
We also exclude stars that appear in the 
compilation of nonmembers from \citet{cab08a} as well as source 668 
from \citet{her07}, which is a galaxy according to spectroscopy from 
\citet{cab08b}.
S~Ori~47 has been classified as a field dwarf rather than a cluster member
based on gravity-sensitive absorption lines \citep{mc04}, but we retain it
in our compilation for the purposes of comparing our IRAC measurements to
those from \citet{cab07} and \citet{sch08}. 
We add stars that have been spectroscopically
confirmed as members by \citet{sch04}, \citet{ken05}, \citet{cab06}, and 
\citet{sac08}, the young stars associated with HH~446, Haro 5-22, and 
Haro 5-35, and $\sigma$~Ori~C and D. We also treat as members
S~Ori~J053946.5$-$022423 and S~Ori~J053912.8$-$022453,
which were identified as candidate members by \citet{bej01} and 
exhibit evidence of disks in their IRAC colors. 
We present our IRAC measurements for this compilation of probable members in 
Table~\ref{tab:mem}, which includes the source identifications 
from the 2MASS Point Source Catalog, \citet{her07}, \citet{bej99,bej01}, 
and the Henry Draper (HD) Catalog. 
We note that the evidence of membership varies significantly among 
the sources in Table~\ref{tab:mem}, and that some of them are better
characterized as possible members rather than probable members 
\citep[e.g., S~Ori~70,][]{bur04}.

\subsection{Candidate Members with IRAC Excesses}
\label{sec:cand}

\begin{figure}
\plotone{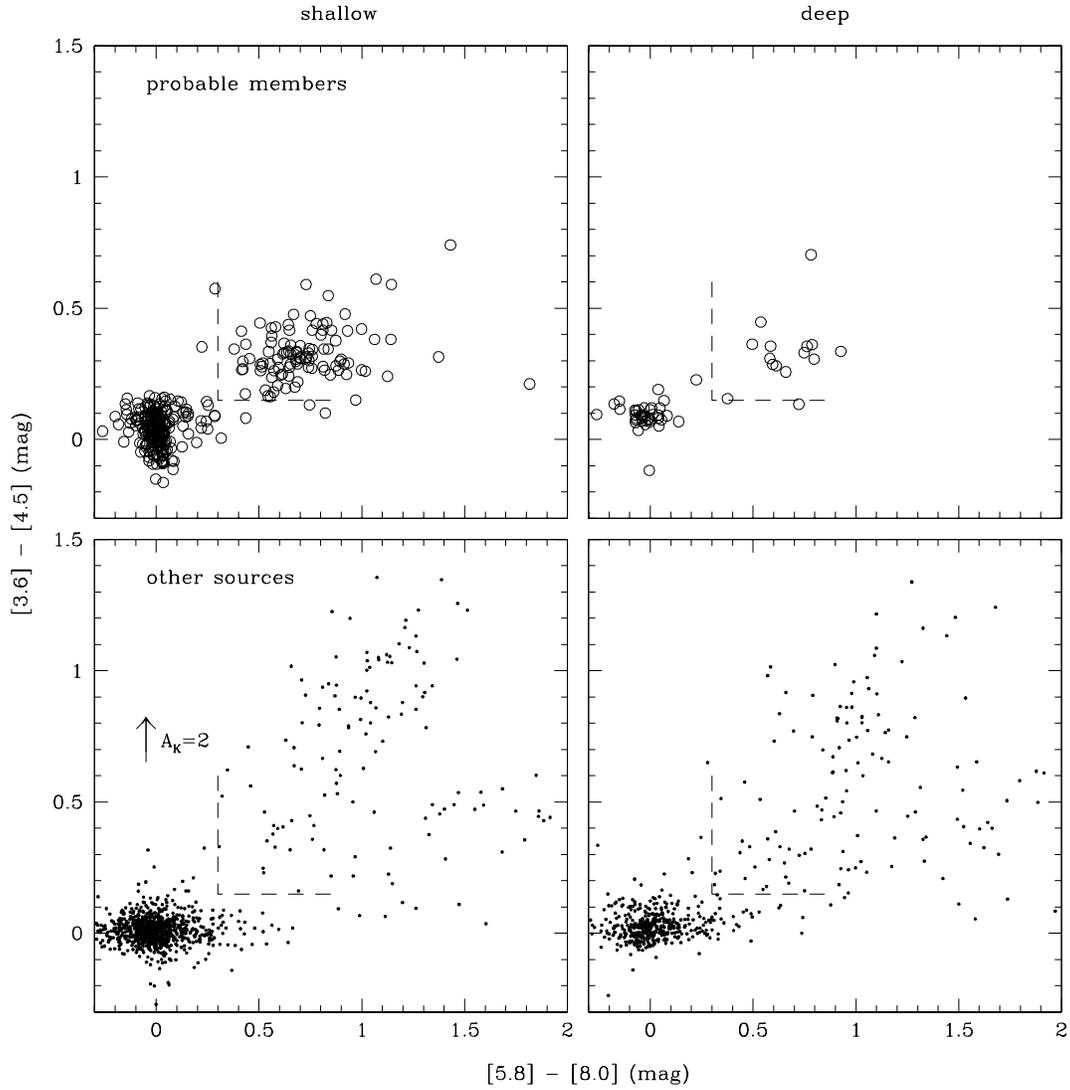}
\caption{
{\it Spitzer} IRAC color-color diagrams for the shallow and deep images of the 
$\sigma$~Ori cluster ({\it left and right}). 
{\it Top}: Probable members of the cluster ({\it circles}) exhibit either 
neutral colors consistent with stellar photospheres or significantly redder 
colors that are indicative of circumstellar disks.
{\it Bottom}: Among the other point sources in the IRAC data ({\it points}),
candidate disk-bearing members of the cluster can be identified 
through their red colors ($[5.8]-[8.0]>0.3$, $[3.6]-[4.5]>0.15$, 
{\it dashed line}). Most of these candidates are probably galaxies based
on their colors and magnitudes (see Fig.~\ref{fig:14}).
Only objects with photometric errors less than 0.1~mag in all four bands
are shown in these diagrams.
The reddening vector is based on the extinction law from \citet{fla07}.
}
\label{fig:1234}
\end{figure}

\begin{figure}
\plotone{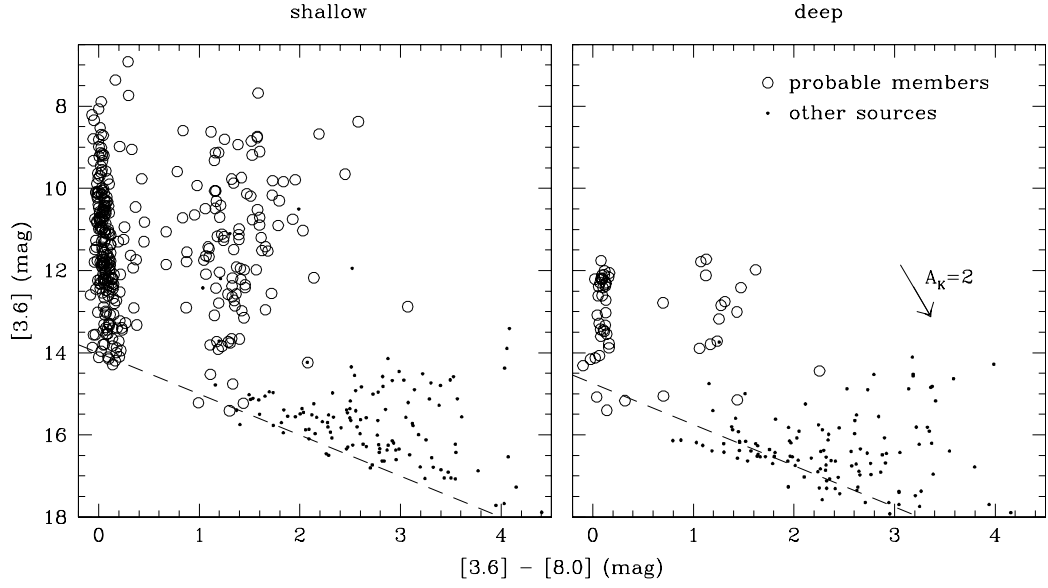}
\caption{
{\it Spitzer} color-magnitude diagrams for the shallow and deep 
images of the $\sigma$~Ori cluster ({\it left and right}).
We show probable members of the cluster ({\it circles}) and sources from 
Figure~\ref{fig:1234} 
that have red colors ($[3.6]-[4.5]>0.15$, $[4.5]-[5.8]>0.15$, 
$[5.8]-[8.0]>0.3$) and lack membership data ({\it points}). 
Most of the latter sources ere probably galaxies based on their faint 
magnitudes and very red colors. Only objects with photometric errors less 
than 0.1~mag in all four bands are shown in these diagrams.
The 8~\micron\ completeness limits are indicated ({\it dashed lines}).
The reddening vector is based on the extinction law from \citet{fla07}.
}
\label{fig:14}
\end{figure}

A color-color diagram constructed from IRAC data is a convenient tool for 
efficiently identifying stars that exhibit IR excess emission indicative 
of disks. In Figure~\ref{fig:1234}, we plot diagrams of $[3.6]-[4.5]$ versus
$[5.8]-[8.0]$ for the probable members of $\sigma$~Ori that are not saturated 
and that have photometric uncertainties less than 0.1~mag in all of the 
IRAC bands. The data from the shallow and deep images are shown separately.
The stars in these diagrams reside near the origin or have significantly
redder colors. These distinct populations are typical of young stellar 
populations \citep{har05} and represent stellar photospheres and stars with 
disks, respectively. In previous IRAC surveys of Taurus, Lupus, and Chamaeleon 
\citep{luh06tau2,luh08cha,all07}, we have used color criteria of 
$[3.6]-[4.5]>0.15$ and $[5.8]-[8.0]>0.3$ to identify stars that might be
new disk-bearing members. As shown in Figure~\ref{fig:1234}, these colors
also encompass most of the red population of probable members of $\sigma$~Ori. 
Therefore, we have applied these criteria to the data in the bottom panels of 
Figure~\ref{fig:1234} to search for new members with disks. 
To further refine the sample of candidates, we have adopted an additional
criterion of $[4.5]-[5.8]>0.15$, which is satisfied by most stars with disks
in $\sigma$~Ori and in other clusters. 
The resulting candidates are plotted in the IRAC color-magnitude diagrams 
in Figure~\ref{fig:14}. 
Most of these sources are redder and fainter than the probable members of 
$\sigma$~Ori, and thus are likely to be galaxies.

Seven of the candidates are in the list of uncertain members from 
\citet{her07}, consisting of sources 336, 361, 395, 614, 633, 916, 
and 950 from that study. Based on their faint magnitudes and red colors, 
sources 336, 361, 916, and 950 are either low-mass protostars or galaxies. 
Indeed, 950 is extended in $K$-band images from UKIDSS, indicating that 
it is probably a galaxy.  Among the remaining candidates that are not in 
\citet{her07}, we have searched for sources that have 
colors similar to those of the probable members ($[3.6]-[8.0]<2$), 
that are point sources in IRAC and UKIDSS images, and that are
detected in the $Z$ and $K$ bands by UKIDSS. 
The five most promising candidate members are presented in Table~\ref{tab:cand}.
In addition to our IRAC data, we include photometry for these candidates
measured from the 24~\micron\ images from \citet{her07}.
Two of the candidates, 2MASS~J05383446$-$0253514 and 2MASS~J05375398$-$0249545,
are below the cluster sequence in the diagram of $V$ versus $V-J$ from 
\citet{her07}, which may indicate that they are seen in scattered light
(e.g., edge-on disks) if they are bona fide members.
The latter candidate also appears near the lower edge of the cluster sequence
in $V$ versus $V-I$ \citep{she04}.
Although it is underluminous in $V$ versus $V-J$, 
2MASS~J05383446$-$0253514 does appear within the cluster sequence in
$I$ versus $I-K$ \citep{cab08a}, probably because of $K$-band excess emission.
2MASS~J05375398$-$0249545 was detected in an H$\alpha$ objective prism 
survey \citep{wea04}, providing additional evidence that it could be
a young star. Based on their IRAC magnitudes, these two candidates should 
have masses of $\sim0.5$~$M_\odot$ if they are cluster members. 
The candidates IRAC J05391066$-$0229238 and IRAC J05384459$-$0225433
have much fainter magnitudes that are indicative of brown dwarfs.
The former underwent significant variability in all IRAC bands between
the shallow and deep images.
The final candidate, IRAC J05384729$-$0235194, exhibits very red IRAC colors
that are consistent with either a protostar or a galaxy. 
This object is only $6\arcsec$ from source 726 from \citet{her07} and 
is $1\arcmin$ from $\sigma$~Ori.  Its close proximity to the center of the
cluster and its relatively bright magnitude compared to most galaxies 
(see Figure~\ref{fig:14}) tend to favor membership.

\section{Disk Population}
\label{sec:pop}

\subsection{Identifying the Members that have Disks}
\label{sec:class}

\begin{figure}
\plotone{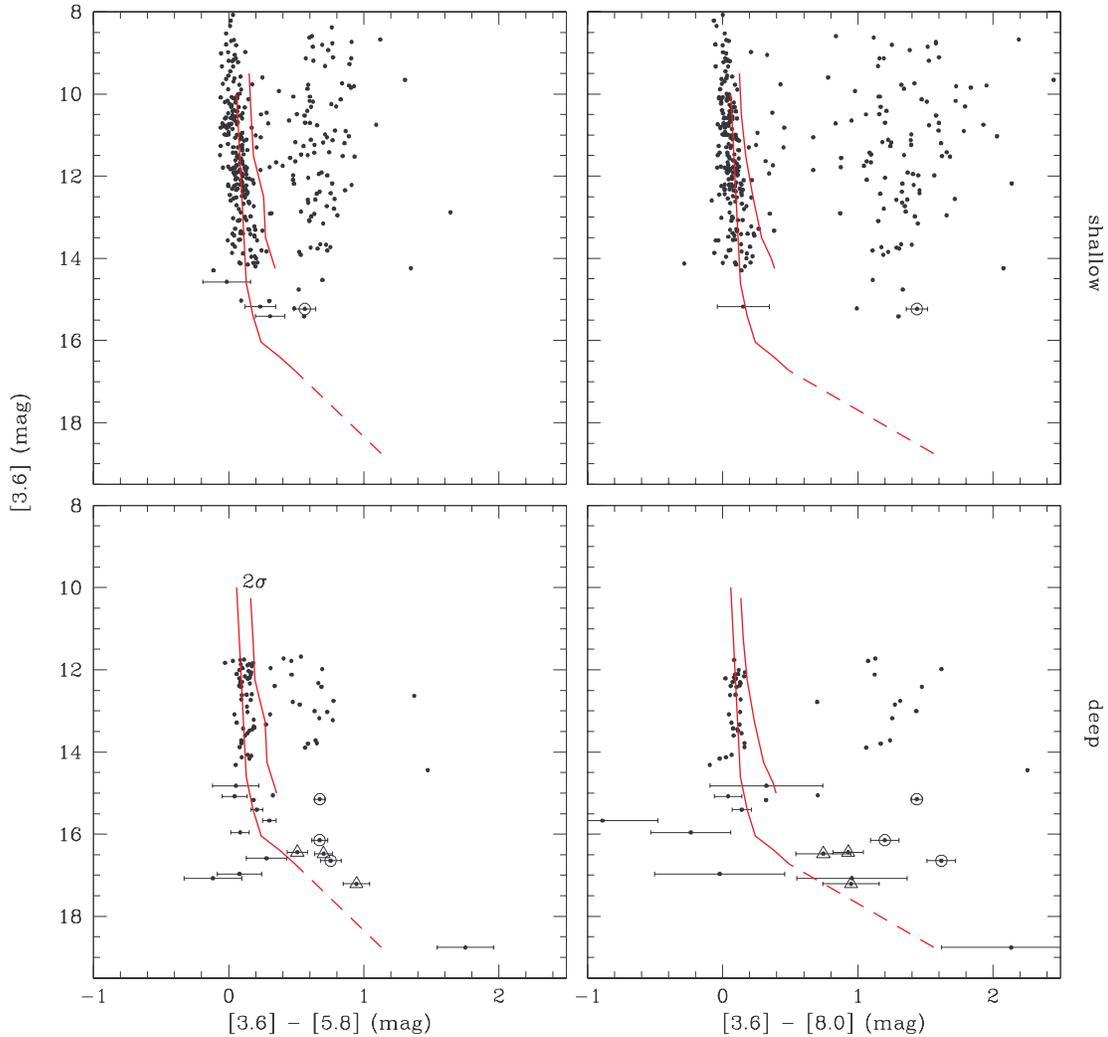}
\caption{
{\it Spitzer} color-magnitude diagrams for probable members of $\sigma$~Ori 
in the shallow and deep images of the cluster ({\it top and bottom}).
For comparison, we show
a fit to the sequence of diskless members of Chamaeleon~I after adjusting 
it to the distance of $\sigma$~Ori ({\it left solid lines}). This fit is 
connected to the typical colors of T6 dwarfs for the magnitude of the T dwarf
S~Ori~70 ({\it dashed lines}). The $2\sigma$ scatter of the $\sigma$ Ori
sequence about this fit represents out adopted threshold for identifying
sources that exhibit significant color excesses at brighter magnitudes
($[3.6]<14.5$, {\it right solid lines}). At the fainter levels encompassing
the brown dwarf candidates that were analyzed by \citet{zap07} 
and \citet{sch08}, we have identified color excesses by considering
the Poisson uncertainties in the colors ({\it error bars}), the size of the 
scatter in colors, and the general tendency to overestimate fluxes for
sources with very low SNRs (\S~\ref{sec:class}).
The brown dwarf candidates that we classify as ``disk?" or ``disk" in 
Table~\ref{tab:class} are indicated ({\it triangles and circles}).
}
\label{fig:d14}
\end{figure}

To characterize the disk population among low-mass members of the $\sigma$~Ori
cluster, we begin by using the IRAC data to identify the probable members
that are likely to have disks. 
Because a disk produces greater excess emission above a stellar photosphere
at longer wavelengths, we select the 5.8 and 8.0~\micron\ bands of IRAC
for measuring disk emission. To construct colors with these data that
can be used for measuring excess emission, we combine them with the IRAC 
data at 3.6~\micron, which are available for nearly all of the probable 
members that are measured at 5.8 and 8.0~\micron. 
Measuring excess emission from $[3.6]-[5.8]$
and $[3.6]-[8.0]$ requires estimates of the intrinsic colors for stellar 
photospheres and their dependence on spectral type. 
We can obtain these estimates by plotting the colors as a function of 
magnitude, which is a reasonable proxy for spectral type in a sample of stars
that are roughly coeval and have low reddening ($A_V<5$). 
In Figure~\ref{fig:d14}, we show color-magnitude diagrams of this kind for 
the probable members of $\sigma$~Ori. The stars that lack disk emission
in the IRAC bands form a well-defined sequence near colors of zero.
In both $[3.6]-[5.8]$ and $[3.6]-[8.0]$, 
the sequence becomes slightly redder with fainter magnitudes,
which reflects the dependence on spectral type. For comparison, we include
in Figure~\ref{fig:d14} a fit to the sequence of diskless members of 
the Chamaeleon~I star forming region \citep{luh08cha}.
The averages of $[3.6]-[5.8]$ and $[3.6]-[8.0]$ for all stars detected
by IRAC toward Chamaeleon~I, which are mostly background stars, are 0.05~mag
greater than the average values in $\sigma$~Ori, which is consistent with
the difference in average extinction for the two regions.
Therefore, we have dereddened the sequence for Chamaeleon~I accordingly
in each color in Figure~\ref{fig:d14}. We have also corrected for differences
in ages and distances by adding 1.6~mag to the Chamaeleon~I fits, which 
is the average difference between the two cluster sequences in 
[3.6] versus spectral type. After applying these color and magnitude
offsets, the sequence for Chamaeleon~I agrees well that of $\sigma$~Ori.

To identify members that exhibit significant excess emission in
$[3.6]-[5.8]$ and $[3.6]-[8.0]$ and thus are likely to have disks, 
we must properly account for errors and biases in the photometric 
measurements, particularly near the detection limits of the data.
As mentioned in \S~\ref{sec:obs}, our reported errors 
include Poisson errors and a 2\% uncertainty in the calibration of IRAC.
However, additional systematic errors are also present, such as
location-dependent variations in the calibration. The precision with which
we can measure excess emission is also limited by the intrinsic scatter
of photospheric colors among cluster members at a given magnitude. 
Fortunately, in a well-populated sample of stars, the sum of these 
uncertainties is reflected in the spread in colors at a given magnitude 
among diskless cluster members. 
Therefore, we can use the widths of the sequences in 
Figure~\ref{fig:d14} to define color thresholds for identifying sources that 
have significant color excesses.
For the shallow data, we have computed the standard deviation of colors in 
each sequence as a function of magnitude for the range of magnitudes in
which the sequences are sufficiently populated ($[3.6]<14.25$). 
Stars that are redder than the 2~$\sigma$ thresholds are classified as having
disks. Because the sequences in the deep images contain fewer stars, 
we adopt for those data the thresholds from the shallow images after
shifting them to fainter magnitudes by 0.75~mag, which is the difference 
in completeness limits between the shallow and deep [8.0] data. 
The classifications produced by $[3.6]-[5.8]$ and $[3.6]-[8.0]$ agree for
99.5\% of the sources. For the remaining 0.5\% of the sample, we adopt the 
classifications
from $[3.6]-[8.0]$ since this color is more sensitive to disk emission.
Our classifications apply only to inner disks that are capable of producing 
emission at 8~\micron. A few of the stars that exhibit photospheric
colors in the IRAC bands may have disks with inner holes that produce
excess emission only at longer wavelengths.

At the faintest magnitudes in Figure~\ref{fig:d14}, the photospheric sequences 
are too sparsely populated for reliable measurements of their widths.
To classify sources that are fainter than the 2~$\sigma$ thresholds, 
we rely on a comparison
of the Chamaeleon~I sequence to the Poisson errors, which are indicated 
in Figure~\ref{fig:d14} for the brown dwarf candidates that were considered 
by \citet{zap07} and \citet{sch08}. For the intrinsic photospheric colors 
of the faintest source, S~Ori~70, we adopt the average colors measured
for field dwarfs \citep{pat06} that have a similar spectral type 
\citep[T6;][]{bur06}.
At the faintest magnitudes, the Poisson errors should be much larger 
than the systematic errors from the location-dependent variations in the
calibration and the intrinsic scatter in photospheric colors. 
However, there remains an additional source of error that must be considered.
In background-limited images, a faint source is preferentially detected 
if upward fluctuations of the background noise bring it above the detection 
limit. As a result, the fluxes measured for sources with very low
signal-to-noise ratios (SNRs) are systematically overestimated. 
For instance, fluxes near the detection limits of the Two-Micron 
All-Sky Survey \citep[2MASS;][]{skr06} are overestimated by an average
of $\sim0.4$~mag \citep{bei03}.
At SNR$<5$, this bias and the Poisson errors are comparable to color excesses
from disks, making such measurements difficult to use in our analysis.
By comparing the color excesses relative to the Chamaeleon~I
sequence to the sizes of the Poisson errors and ignoring the
data at SNR$<5$, we have identified the faint sources in Figure~\ref{fig:d14}
that are likely to have disks. 
These classifications for the brown dwarf candidates considered by 
\citet{zap07} and \citet{sch08} are presented in Table~\ref{tab:class}. 
The available data at 5.8 and 8.0~\micron\ offer insufficient SNRs for 
classifying six of the sources in Table~\ref{tab:class} as well as 
S~Ori~J053844.5$-$025512, S~Ori~J053932.4$-$025220, S~Ori~J053929.4$-$024636,
S~Ori~J053944.5$-$025959, and S~Ori~J054007.0$-$023604. 

Based on our classifications in Table~\ref{tab:class}, S~Ori~60 is the 
faintest object in $\sigma$~Ori that exhibits significant evidence of a disk.
If we adopt a distance modulus of 7.65 (\S~\ref{sec:fraction}) and photometry
from UKIDSS\footnote{We adopt the UKIDSS magnitudes measured with an aperture
radius of $1\arcsec$ \citep{dye06}.},
then S~Ori~60 has $M_J=11.3$, $M_H=10.5$, and $M_K=9.9$.
In comparison, the least luminous disk-bearing brown dwarf found in other
young clusters is Cha~J11070768$-$7626326, which has slightly fainter 
magnitudes of $M_J=11.56$, $M_H=10.75$, and $M_K=9.86$ \citep{luh08cha}. 
The faintest $\sigma$~Ori source that has marginal excess emission is S~Ori~66.
It is $\sim0.8$~mag fainter than S~Ori~60 according to UKIDSS.

\subsection{Disk Fraction}
\label{sec:fraction}

We now use the IRAC classifications derived in the previous section 
to compute the disk fraction among the probable members of $\sigma$~Ori.
We wish to measure the dependence of this disk fraction on stellar mass.  
Ideally, the masses would be estimated by combining spectral types and 
luminosities with theoretical evolutionary models. However, spectral 
classifications are unavailable for many of the probable members of 
$\sigma$~Ori. Therefore, we adopt $M_J$ as a proxy for stellar mass, as done 
by \citet{her07}. We adopt the $J$ measurements from 2MASS and UKIDSS when 
possible. For the few sources that do not have photometric errors less
than 0.2~mag in these surveys,
we use the $J$ data from \citet{bej01}, \citet{mar01}, \citet{cab07},
and \citet{zap08}. We assume that the probable members of $\sigma$~Ori 
have negligible extinctions at $J$.
For comparison to $\sigma$~Ori, we also consider other clusters in which 
disk fractions that have been measured for both stars and brown dwarfs using 
IRAC data.
Therefore, we have recomputed the disk fractions for IC~348 and Chamaeleon~I
from \citet{mue07} and \citet{luh08cha} as a function of $M_J$ using the
$J$-band data from 2MASS, \citet{luh03ic}, and \citet{luh07cha} and
extinction estimates from \citet{mue07} and \citet{luh07cha}. 
We adopt distance moduli of 7.5 and 6.05 for IC~348 and Chamaeleon~I,
respectively \citep{her98,luh08blue}. 
To enable a meaningful comparison of the disk fractions of these clusters,
$M_J$ must be corrected for differences in age so that a given value of
$M_J$ corresponds to the same mass in each cluster. For the adopted distances, 
IC~348 and Chamaeleon~I do have similar ages \citep{luh07cha}. 
In order to compare the disk fraction in $\sigma$~Ori to those of the other two
clusters in a manner that is as age-independent as possible, we
adopt a distance modulus that is the sum
of the distance modulus for Chamaeleon~I and the difference of the apparent
magnitudes as a function of spectral type between Chamaeleon~I and 
$\sigma$~Ori (\S~\ref{sec:class}), corresponding to a value of 7.65 (340~pc).
In other words, we are adopting a distance for $\sigma$~Ori that gives
it the same age as IC~348 and Chamaeleon~I. Similar distances have been
adopted in some previous studies of $\sigma$~Ori \citep{bej01,oli02}
while values near 440~pc have been used elsewhere \citep{she04,her07}.
The larger distances would imply that $\sigma$~Ori is younger than
IC~348 and Chamaeleon~I.

Six sources in $\sigma$~Ori are beyond the faint magnitude limit that we 
have selected for plotting the disk fractions ($M_J>12$, $J>19.65$), 
four of which are too faint for a useful constraint on excess emission 
at 5.8 or 8.0~\micron. In the faintest bin of the disk fraction 
for $\sigma$~Ori ($M_J=10$--12), two objects have tentative detections of 
excesses and two others are too faint for classifications. For the purposes 
of this work, we assume that the disks are present in the former but not 
the latter. 

\begin{figure}
\epsscale{0.45}
\plotone{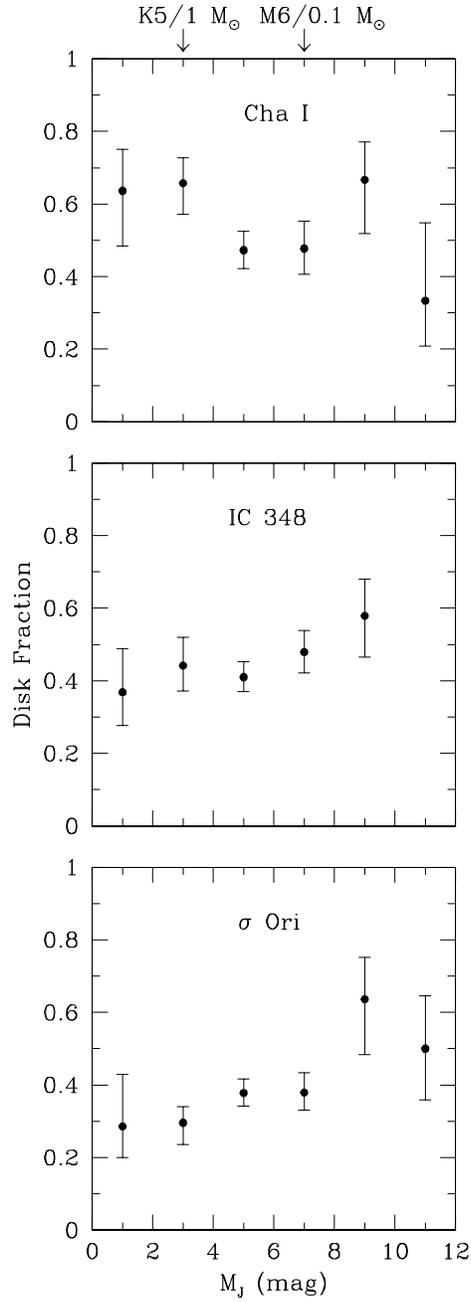}
\caption{
Disk fraction as a function of absolute $J$-band magnitude for probable members 
of the $\sigma$~Ori cluster based on the IRAC color excesses in 
Figure~\ref{fig:d14}. For comparison, we include the disk fractions measured
from IRAC data in IC~348 and Chamaeleon~I \citep{luh05frac,luh08cha,lada06}.
The typical magnitudes for spectral types of K5 ($\sim1$~$M_\odot$) and M6
($\sim0.1$~$M_\odot$) are indicated.
}
\label{fig:diskfraction}
\end{figure}

The disk fractions versus $M_J$ for $\sigma$~Ori, IC~348, and Chamaeleon~I
are listed in Table~\ref{tab:diskfraction} and are plotted in 
Figure~\ref{fig:diskfraction}.
For $M_J>4$ ($M\lesssim0.5$~$M_\odot$), the three regions exhibit similar disk
fractions. In each case, $\sim40$\% of low-mass stars (0.08--0.5~$M_\odot$)
and $\sim50$--60\% of brown dwarfs (0.01--0.08~$M_\odot$) have disks. 
Meanwhile, the disk fractions at $M_J<4$ 
range from $\sim30$\% in $\sigma$~Ori to $\sim65$\% in Chamaeleon~I.
Since these clusters have similar ages, these data suggest that 
the lifetimes of disks around solar-mass stars increase from
$\sigma$~Ori to IC~348 to Chamaeleon~I. 

We note a few caveats regarding the disk fractions in 
Figure~\ref{fig:diskfraction}. A given $M_J$ magnitude corresponds to a larger 
range of stellar masses than a given spectral type. In addition, the near-IR
magnitudes of stars with edge-on disks are typically much fainter than most
other cluster members at the same spectral type and color. As a result,
a star with an edge-on disk can appear at a magnitude bin corresponding to 
substellar masses in a plot of disk fraction versus $M_J$. 
Meanwhile, stars with disks can have systematically brighter $J$ magnitudes
than diskless stars because of disk emission in this band, which would 
cause these sources to appear in magnitude bins that are too bright. 
Because of these effects, the disk fractions versus $M_J$ in 
Figure~\ref{fig:diskfraction} are smoothed versions of the disk fractions
versus mass or spectral type. For instance, the difference between the disk 
fractions of Chamaeleon~I and IC~348 at high masses is more pronounced when 
they are plotted as a function of spectral type or mass \citep{luh08cha}. 
Finally, unlike the samples for IC~348 and Chamaeleon~I, some of the 
sources included in the disk fraction for $\sigma$~Ori lack definitive 
evidence of membership. Thus, our estimate of the disk fraction 
for $\sigma$~Ori is may be underestimated. 
Indeed, \citet{sac08} obtained spectra of 23\% of the probable members 
identified by \citet{her07} and found a field star contamination rate of 20\%.
Because of the spectral classifications that are now available from 
\citet{sac08}, the contamination of our current sample of probable members
should be lower.

\subsection{Comparison to Previous Studies}

\begin{figure}
\epsscale{1}
\plotone{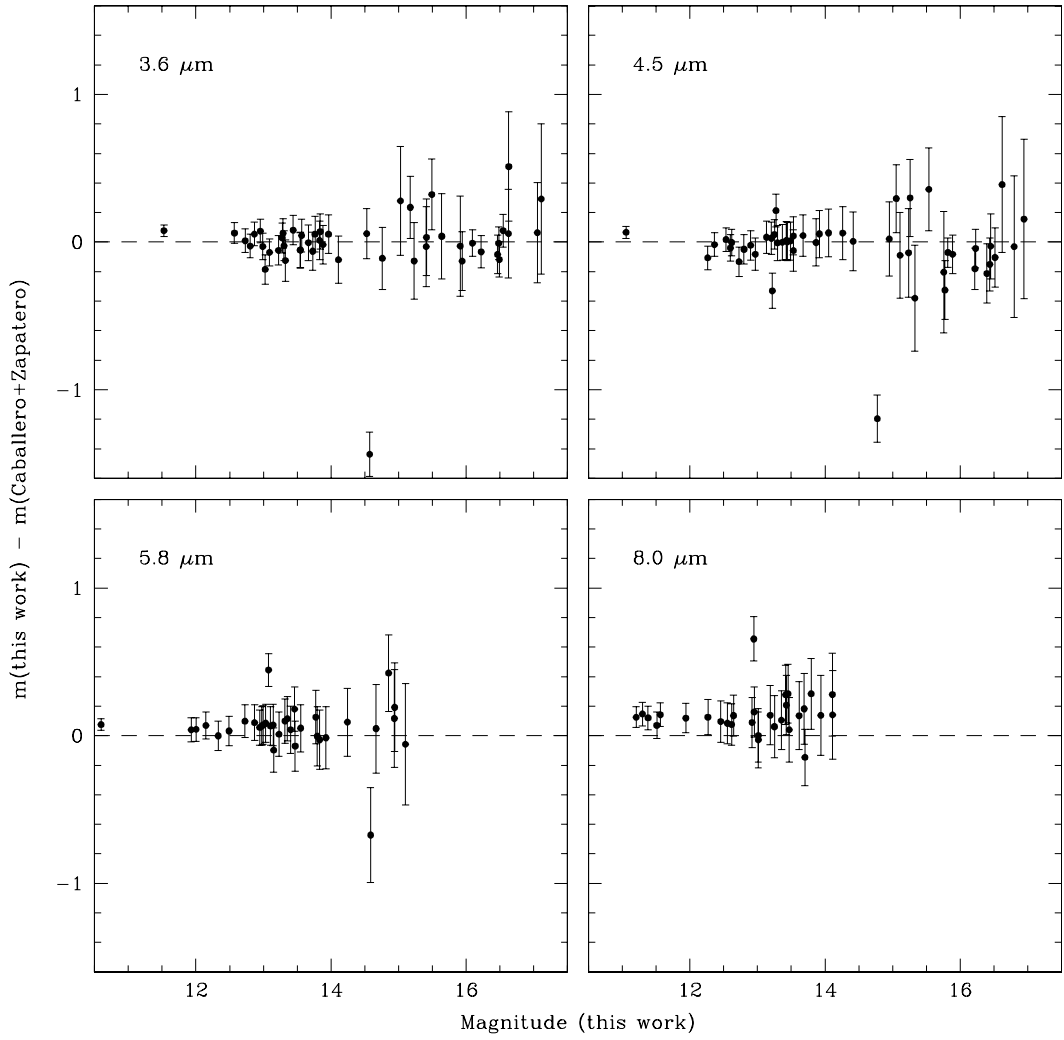}
\caption{
Comparison of the IRAC magnitudes from \citet{cab07} and \citet{zap07}
to our measurements of 
the same objects. Both sets of data were measured from the shallow images
of the $\sigma$~Ori cluster.
The errors reported by \citet{cab07} and \citet{zap07} are indicated.
The formal errors for our measurements range between 0.02--0.12~mag.
}
\label{fig:cab}
\end{figure}

\begin{figure}
\epsscale{1}
\plotone{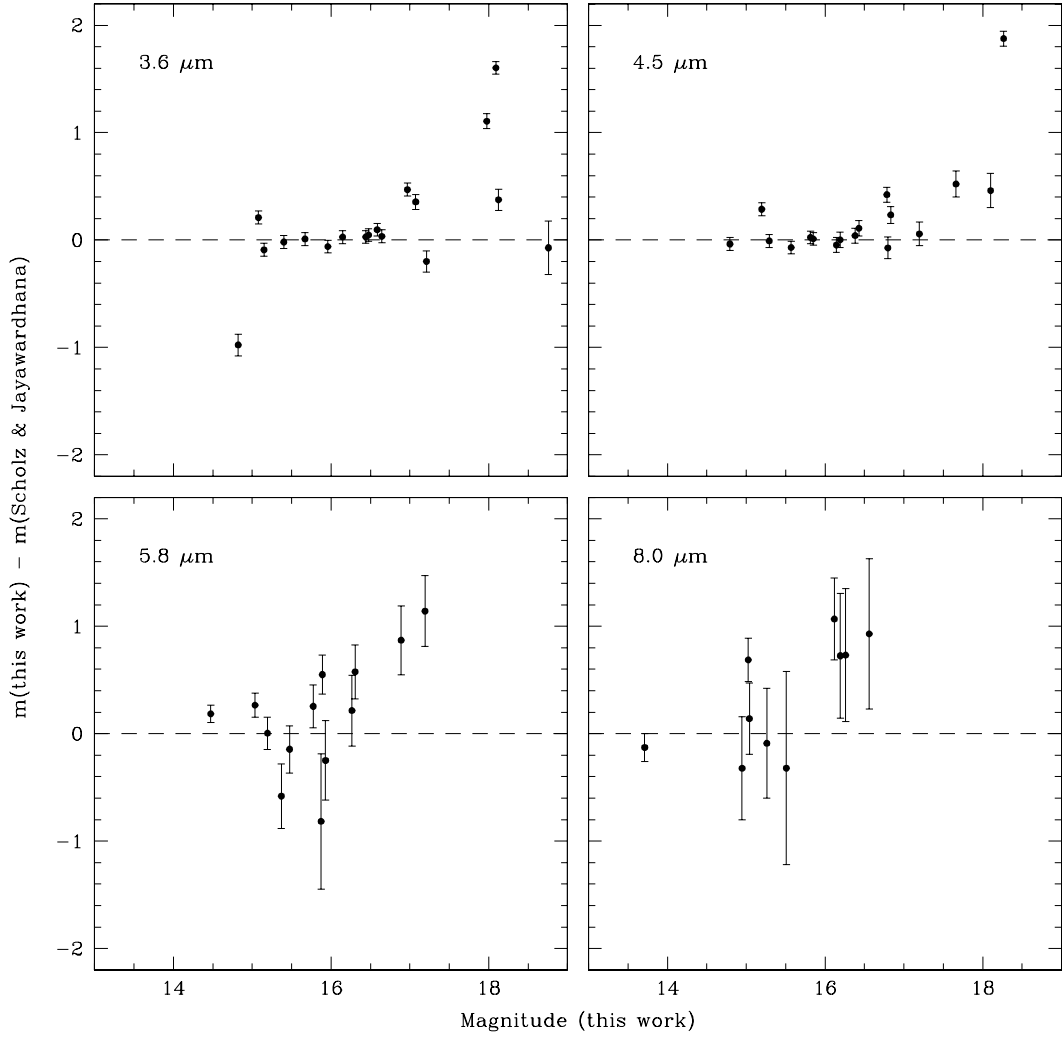}
\caption{
Comparison of the IRAC magnitudes from \citet{sch08} to our measurements of 
the same objects. Both sets of data were measured from the deep images
of the $\sigma$~Ori cluster.
The errors reported by \citet{sch08} are indicated.
The formal errors for our measurements are smaller by factors of 2--3 in most
cases.
}
\label{fig:sch}
\end{figure}

\begin{figure}
\epsscale{1}
\plotone{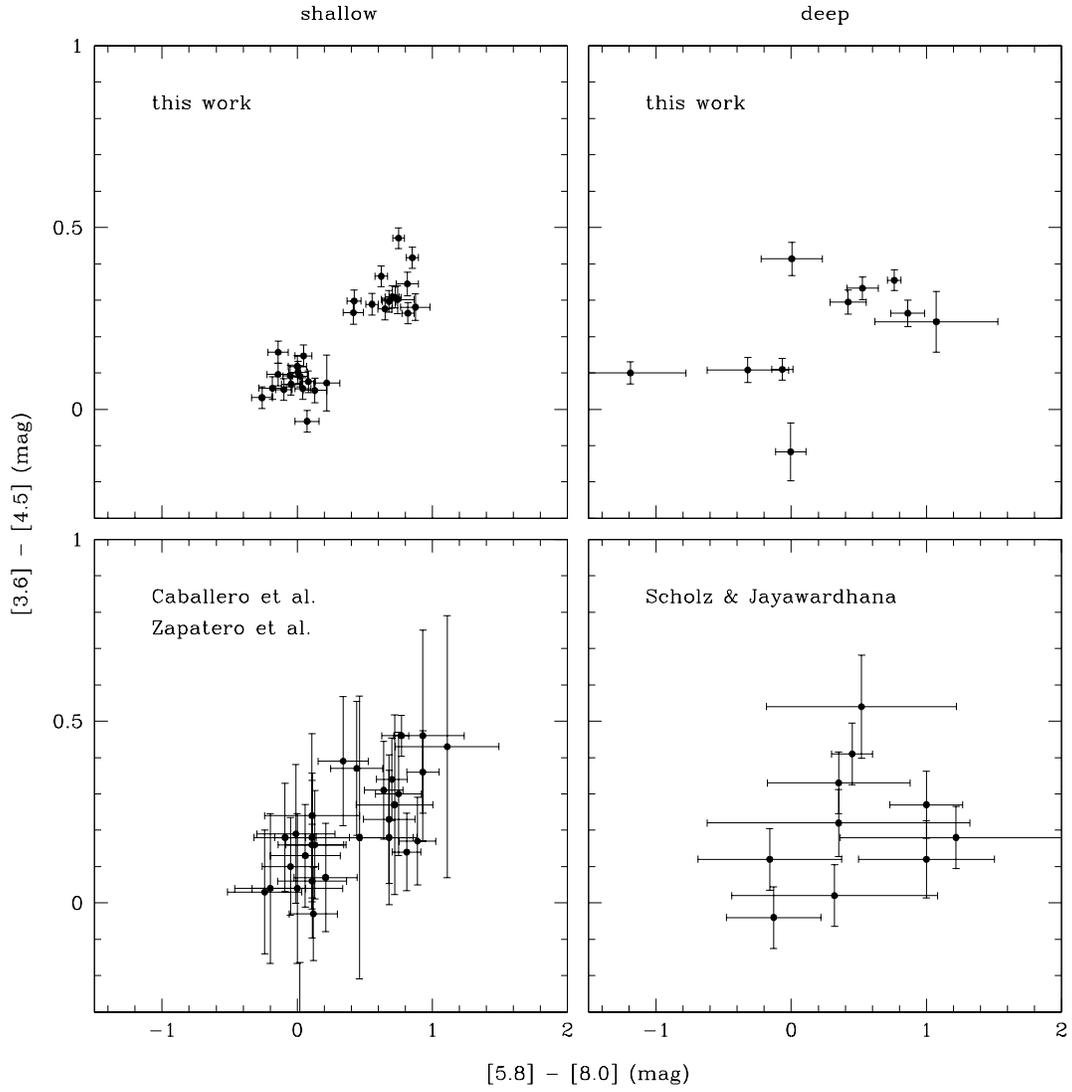}
\caption{
Comparison of the IRAC colors from \citet{cab07}, \citet{zap07}, and 
\citet{sch08} to our measurements of the same objects from the same 
images considered in those studies, which are the shallow and deep images
of the $\sigma$~Ori cluster ({\it left and right}).
}
\label{fig:cabsch}
\end{figure}

As discussed in \S~\ref{sec:intro}, five previous studies have used 
IRAC data to investigate the disk population among brown dwarfs in the
$\sigma$~Ori cluster \citep{her07,cab07,zap07,zap08,sch08}.
To compare the results from those studies and this work, 
we begin by plotting the differences between the previously reported 
IRAC magnitudes and our measurements for the shallow and deep images 
in Figures~\ref{fig:cab} and \ref{fig:sch}, respectively.
The differences in photometry are more than 0.1~mag in many cases 
and reach as high as 1--2~mag. Most of the measurements from \citet{cab07}
and \citet{zap07} agree with our data within their quoted errors, but not 
within our errors. Roughly half of the measurements in common between this work 
and \citet{sch08} differ by an amount that is larger than the errors from 
either study. Our formal errors are much lower than the ones reported 
by \citet{cab07}, \citet{zap07}, and \citet{sch08}, but this in itself does 
not demonstrate that our measurements are truly more accurate.
To test the relative accuracies of these sets of photometry,
we compare them in color-color diagrams in Figure~\ref{fig:cabsch}.
For both the shallow and deep images, our colors are more tightly 
clustered into two groups (diskless and disk-bearing stars) 
than the colors in the other studies, which suggests that our colors indeed 
have lower errors. A similar result was found by \citet{luh08cha} in a 
comparison of their IRAC data to measurements from \citet{dam07} in 
Chamaeleon~I.

\begin{figure}
\epsscale{.8}
\plotone{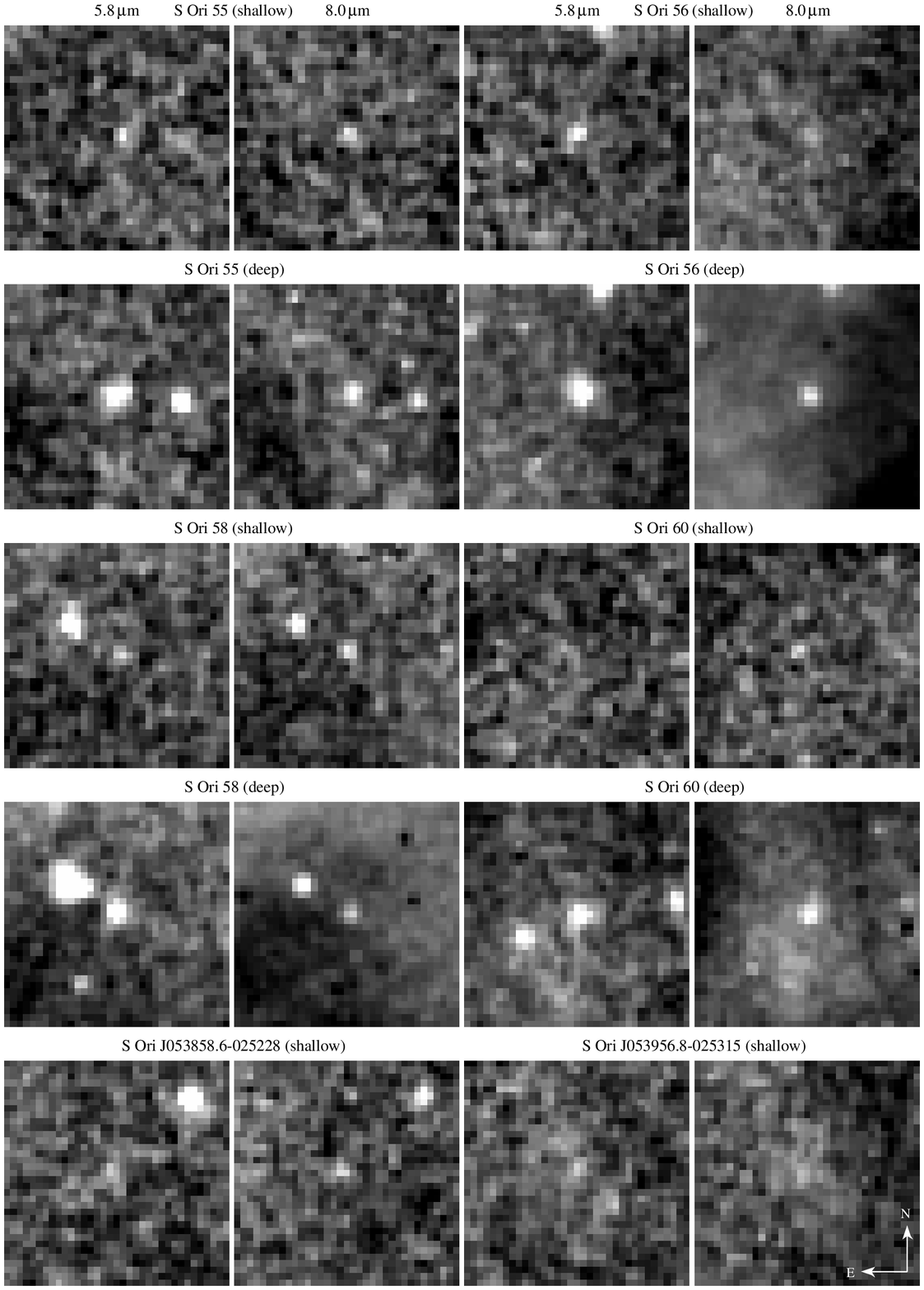}
\caption{
IRAC images at 5.8 and 8.0~\micron\ ({\it odd and even columns}) centered 
on the positions of six of the candidate brown dwarfs 
in $\sigma$~Ori that were discussed by \citet{zap07}. 
Based on the shallow images, \citet{zap07} reported detections of excess 
emission that indicated the presence of disks for all of these sources. 
However, we find that the SNRs of these data are insufficient for reliably 
detecting excess emission. Deep images do show significant excesses for 
S~Ori~56 and S~Ori~60 
and marginal excesses for S~Ori~55 and S~Ori~58 (Figure~\ref{fig:d14}).
Each image has a size of $30\arcsec\times30\arcsec$ and is displayed linearly 
from $F-2$~$\sigma$ to $F+5$~$\sigma$, where $F$ and $\sigma$ are the median 
and standard deviation of the background emission, respectively.  
}
\label{fig:image1}
\end{figure}

We now compare our classifications of the IRAC data in $\sigma$~Ori
to those derived in previous work.
The classifications from \citet{cab07} and our earlier survey of 
$\sigma$~Ori \citep{her07} agreed for the 25 sources that were considered 
by both studies, which were predominantly low-mass stars.
The samples from \citet{zap07} and \citet{sch08} consisted of fainter brown
dwarf candidates. These sources are listed in Table~\ref{tab:class},
where we compare the classifications from \citet{cab07}, \citet{zap07}, 
\citet{sch08}, and this work.  
The four studies differ significantly in the sources
that are found to exhibit excess emission from disks.
\citet{zap07} reported excess emission at 5.8 or 8.0~\micron\ in the 
shallow images for seven candidate brown dwarfs.  However, we find that 
six of these sources have SNRs in those data that are too low for useful 
photometry, as illustrated in Figure~\ref{fig:image1}. 
Four of these six candidates are within the deep images from \citet{sch08},
which provide much better detections. Using the deep data, we measure
significant excess emission for S~Ori~56 and S~Ori~60 and marginal excesses
for S~Ori~55 and S~Ori~58. \citet{sch08} derived the same classifications
for S~Ori~58 and S~Ori~60, but concluded that the other two sources lack disks.
The seventh object from \citet{zap07}, S~Ori~54, is too close ($5\arcsec$) to 
a red galaxy for an accurate measurement at 8~\micron\ using either the 
shallow or deep images. We do not find excess emission for this source at 
5.8~\micron, where the contrast relative to the galaxy is more favorable. 

\begin{figure}
\epsscale{0.5}
\plotone{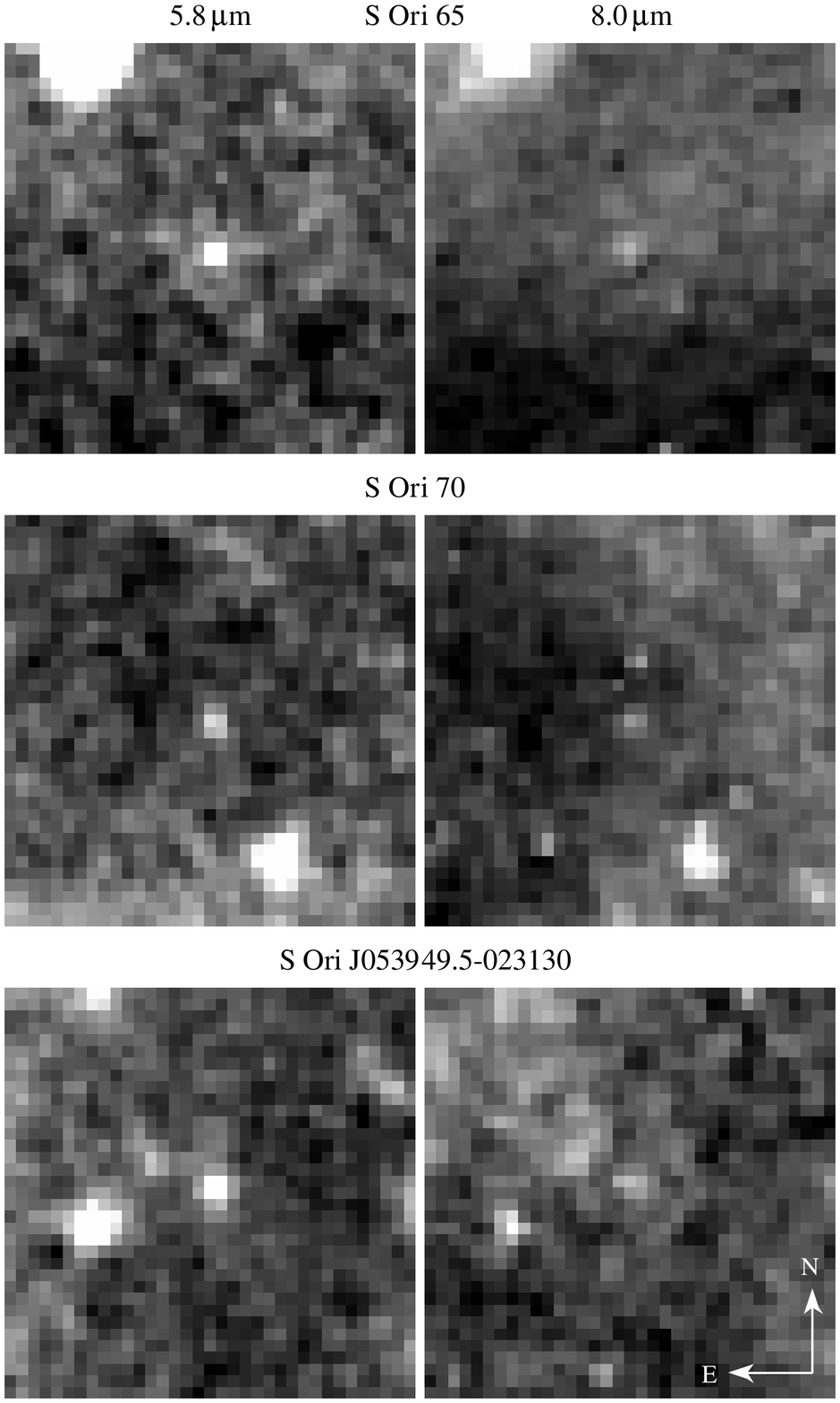}
\caption{
Deep IRAC images at 5.8 and 8.0~\micron\ ({\it left and right}) centered on the 
positions of three of the candidate brown dwarfs 
in $\sigma$~Ori that were discussed by \citet{sch08}. 
Based on these images, \citet{sch08} reported detections of excess emission 
that indicated the presence of disks (their detection was only tentative for 
S~Ori~70). However, we find that the SNRs of the 8~\micron\ data are 
insufficient for reliably detecting excess emission.
In the images at 5.8~\micron, which have higher SNRs, S~Ori~65 and 
S~Ori~J053949.5$-$023130 do not exhibit excesses. 
S~Ori~70 does have an excess at 5.8~\micron, but this measurement 
may be overestimated because of its low SNR \citep{bei03}. 
Each image has a size of $30\arcsec\times30\arcsec$ and is displayed linearly 
from $F-2$~$\sigma$ to $F+5$~$\sigma$, where $F$ and $\sigma$ are the median 
and standard deviation of the background emission, respectively. 
}
\label{fig:image2}
\end{figure}

The three remaining sources Table~\ref{tab:class} that have discrepant 
classifications are S~Ori~65, S~Ori~70, and S~Ori~J053949.5$-$023130. 
\citet{sch08} found significant excess emission for S~Ori~65 and 
S~Ori~J053949.5$-$023130 at 5.8 and 8.0~\micron\ from their deep images.
However, our measurements at 5.8~\micron\ do not exhibit excesses and the 
detections at 8.0~\micron\ are too weak to be useful 
(see Figure~\ref{fig:image2}).
Although S~Ori~65 has $[3.6]-[8.0]\sim1$ in our data, which implies an excess,
the photometry at 8.0~\micron\ has an error of 0.4~mag, corresponding to
SNR$\sim3$. As discussed in \S~\ref{sec:class}, measurements with such
low values of SNR are subject to flux overestimation.
To illustrate that color excesses with such large errors are not
significant, we note that another source with comparable errors at 8.0~\micron, 
S~Ori~50, has a nominal color that is unphysical for stellar sources 
($[3.6]-[8.0]\sim-0.9$). Finally, based on their deep images,
\citet{sch08} suggested that S~Ori~70 may 
have excesses in $[3.6]-[4.5]$ and $[3.6]-[5.8]$ relative to typical colors of 
T6 dwarfs, which they attributed to either a disk or low surface gravity and 
cited as possible evidence of youth and membership in the $\sigma$~Ori cluster.
\citet{zap08} arrived at a similar conclusion through a measurement of
$[3.6]-[4.5]$ from the shallow images.
We measure $[3.6]-[4.5]=1.56\pm0.07$ and $[3.6]-[5.8]=1.75\pm0.21$ from
the deep data. In comparison, the colors of
field dwarfs between T6 and T6.5 range from 0.93 to 1.46 in $[3.6]-[4.5]$ and
from 0.87 to 1.19 in $[3.6]-[5.8]$ \citep{pat06}. 
Thus, we find that S~Ori~70 is only slightly redder in $[3.6]-[4.5]$
than the reddest field dwarfs. Given the low SNR at 5.8~\micron\ (see
Figure~\ref{fig:image2}), the apparent excess in this band could be 
caused by a combination of Poisson errors and flux overestimation.

\section{Conclusions}
\label{sec:conc}

We have investigated the disk population among stars and brown dwarfs in the 
$\sigma$~Ori cluster using mid-IR images obtained with IRAC onboard
the {\it Spitzer Space Telescope}.
For this study, we have employed all available IRAC data for $\sigma$~Ori, 
which consist of shallow images (80.4~s) from \citet{her07} and deep images 
($\sim1100$~s) from \citet{sch08}. 
We measured photometry for all sources detected in these images and
searched the resulting data for new members of the cluster based on 
the red colors that are expected from stars with disks.
The five most promising candidates have colors and magnitudes that are 
suggestive of edge-on disks, brown dwarfs, and a low-mass protostar.
We then examined the IRAC colors for $\sim300$ probable cluster members 
found in previous studies and identified the ones that are likely to have 
disks. In doing so, we have attempted to fully account for the errors
and biases in the photometry of the brown dwarf candidates 
\citep[e.g., flux overestimation at low SNR;][]{bei03}, some of which
are near the detection limits of the IRAC data. 
S~Ori~60 is the faintest candidate member of $\sigma$~Ori that
exhibits significant IR excess emission. This object is comparable
in luminosity to the faintest brown dwarf that shows evidence of a disk 
in other star-forming regions \citep{luh08cha}.
Using our classifications of the IRAC data, we computed the disk fraction 
as a function of $M_J$, which acts as a proxy for stellar mass. 
The disk fractions for low-mass stars (0.08--0.5~$M_\odot$) and brown dwarfs
(0.01--0.08~$M_\odot$) are $\sim40$\% and $\sim60$\%, respectively, which are 
similar to the disk fractions derived from IRAC surveys of two other
clusters near the same age, IC~348 and Chamaeleon~I \citep{mue07,luh08cha}.

Although our disk fraction for brown dwarfs in $\sigma$~Ori
is similar to other estimates based on the IRAC data considered here
\citep{cab07,zap07,sch08}, our photometric measurements and our
classifications of the IRAC colors (disk vs.\ no disk) differ significantly 
from those in previous studies. 
For instance, \citet{zap08} and \citet{sch08} suggested that the T dwarf 
S~Ori~70 may exhibit 
color excesses in $[3.6]-[4.5]$ and $[3.6]-[5.8]$ relative to field 
dwarfs at the same spectral type. They cited these non-standard colors as 
possible evidence of youth, which would indicate that S~Ori~70 is a cluster 
member rather than a field dwarf.  However, we have found that this object is
not significantly redder in $[3.6]-[4.5]$ than field dwarfs and that its SNR 
may be too low at 5.8~\micron\ for a useful measurement of $[3.6]-[5.8]$. 
Therefore, we conclude that the IRAC data do not provide firm constraints
on the membership of S~Ori~70.

\acknowledgements

K.~L. was supported by grant AST-0544588 from the National Science Foundation. 
This publication has made use of NASA's Astrophysics Data System
Bibliographic Services, the SIMBAD database, operated at CDS in Strasbourg,
France, and data products from 2MASS and DSS.
2MASS is a joint project of the University of Massachusetts
and the Infrared Processing and Analysis Center/California Institute
of Technology, funded by the National Aeronautics and Space
Administration and the National Science Foundation.
DSS was produced at the Space Telescope Science Institute under
U.S. Government grant NAG W-2166.

\clearpage

\LongTables

\begin{deluxetable}{lllllll}
\tabletypesize{\scriptsize}
\tablewidth{0pt}
\tablecaption{{\it Spitzer} Photometry for Probable Members of 
$\sigma$ Ori\label{tab:mem}}
\tablehead{
\colhead{ID\tablenotemark{a}} &
\colhead{Name\tablenotemark{b}} &
\colhead{[3.6]} &
\colhead{[4.5]} &
\colhead{[5.8]} &
\colhead{[8.0]} &
\colhead{Date}
}
\startdata
2MASS J05371537$-$0230534 &              \nodata &            sat &            sat & 10.69$\pm$0.03 & 10.79$\pm$0.03 & 2006 Sep 28  \\
2MASS J05371869$-$0240218 &                 SO 9 & 11.01$\pm$0.02 & 10.98$\pm$0.02 & 10.81$\pm$0.03 & 10.94$\pm$0.03 & 2004 Oct 9  \\
                     &                      &            out & 11.01$\pm$0.02 &            out & 10.94$\pm$0.03 & 2006 Sep 28  \\
2MASS J05372306$-$0232465 &                SO 27 & 13.02$\pm$0.02 & 12.95$\pm$0.02 & 12.90$\pm$0.04 & 13.03$\pm$0.04 & 2004 Oct 9  \\
                     &                      & 13.03$\pm$0.02 & 12.95$\pm$0.02 & 12.89$\pm$0.03 & 12.90$\pm$0.03 & 2006 Sep 28  \\
IRAC J05372470$-$0231523 &               S Ori 66 & 17.43$\pm$0.06 & 16.82$\pm$0.08 &        \nodata &        \nodata & 2004 Oct 9  \\
                     &                      & 17.21$\pm$0.03 & 16.80$\pm$0.04 & 16.27$\pm$0.09 & 16.26$\pm$0.21 & 2006 Sep 28  \\
IRAC J05372587$-$0234320 &               S Ori 55 & 16.49$\pm$0.04 & 16.23$\pm$0.05 &        \nodata &        \nodata & 2004 Oct 9  \\
\enddata
\tablecomments{Entries of ``$\cdots$", ``sat", and ``out"
indicate measurements that are absent because of non-detection, saturation,
and a position outside the field of view of the filter, respectively.}
\tablenotetext{a}{2MASS identifications are from the 2MASS Point Source Catalog.
IRAC identifications are based on J2000 coordinates measured with IRAC.}
\tablenotetext{a}{SO identifications are from \citet{her07}.}
\end{deluxetable}

\begin{deluxetable}{llllllll}
\tabletypesize{\scriptsize}
\tablewidth{0pt}
\tablecaption{{\it Spitzer} Photometry for Candidate Members of $\sigma$ 
Ori\label{tab:cand}}
\tablehead{
\colhead{Name} &
\colhead{[3.6]} &
\colhead{[4.5]} &
\colhead{[5.8]} &
\colhead{[8.0]} &
\colhead{Date} &
\colhead{[24]} &
\colhead{Date}
}
\startdata
2MASS J05375398$-$0249545\tablenotemark{a} & 11.10$\pm$0.02 & 10.70$\pm$0.02 & 10.39$\pm$0.03 &   9.80$\pm$0.04 & 2004 Oct 9 & 7.24$\pm$0.04 & 2004 Mar 17 \\
2MASS J05383446$-$0253514\tablenotemark{b} & 10.50$\pm$0.02 & 9.77$\pm$0.02 & 9.14$\pm$0.03 & 8.51$\pm$0.03 & 2004 Oct 9 & 5.68$\pm$0.04 & 2004 Mar 17 \\
 & sat & out & 9.54$\pm$0.03 & out & 2007 Apr 3 & \nodata & \nodata \\
IRAC J05384459$-$0225433 & 15.75$\pm$0.03 & 15.35$\pm$0.03 & 14.96$\pm$0.09 & 14.35$\pm$0.09 & 2004 Oct 9 &  \nodata & \nodata \\
 & 15.87$\pm$0.03 & 15.52$\pm$0.03 & 14.93$\pm$0.04 & 14.48$\pm$0.06 & 2007 Mar 30/Apr 3 & \nodata & \nodata \\
IRAC J05384729$-$0235194 & 14.16$\pm$0.20 & 13.21$\pm$0.12 & 12.72$\pm$0.14 & 11.22$\pm$0.13 & 2007 Apr 3 & \nodata & \nodata \\
IRAC J05391066$-$0229238 & 15.77$\pm$0.03 & 15.15$\pm$0.03 & 14.38$\pm$0.07 & 14.04$\pm$0.06 & 2004 Oct 9 &  \nodata & \nodata \\
 & 16.15$\pm$0.02 & 15.68$\pm$0.02 & 14.89$\pm$0.04 & 14.19$\pm$0.06 & 2007 Apr 3 & \nodata & \nodata \\
\enddata
\tablecomments{Entries of ``sat" and ``out"
indicate measurements that are absent because of saturation
and a position outside the field of view of the filter, respectively.}
\tablenotetext{a}{Identified as a candidate member by \citet{she04}.}
\tablenotetext{b}{Identified as a candidate member by \citet{cab08a}.}
\end{deluxetable}

\begin{deluxetable}{lllll}
\tabletypesize{\scriptsize}
\tablewidth{0pt}
\tablecaption{IRAC Classifications for Candidate Brown Dwarfs in 
$\sigma$ Ori\label{tab:class}}
\tablehead{
\colhead{Name} &
\colhead{C07\tablenotemark{a}} &
\colhead{Z07\tablenotemark{b}} &
\colhead{SJ08\tablenotemark{c}} &
\colhead{this work}} 
\startdata
              S Ori 47\tablenotemark{d} & no disk &     \nodata & no disk & no disk  \\
              S Ori 50 & no disk &    \nodata & no disk & no disk  \\
              S Ori 51 & no disk &    \nodata & no disk & no disk  \\
              S Ori 53 & ? (too faint) &   \nodata & no disk & no disk  \\
              S Ori 54 & \nodata & disk & no disk & no disk  \\
              S Ori 55 & \nodata & disk & no disk & disk?  \\
              S Ori 56 & \nodata & disk & no disk &  disk  \\
              S Ori 58 & \nodata & disk & disk? & disk?  \\
\enddata
\tablenotetext{a}{\citet{cab07}.}
\tablenotetext{b}{\citet{zap07}.}
\tablenotetext{c}{\citet{sch08}.}
\tablenotetext{d}{Spectroscopically classified as a field dwarf rather 
than a cluster member \citep{mc04}.}
\end{deluxetable}

\begin{deluxetable}{lccc}
\tabletypesize{\scriptsize}
\tablewidth{0pt}
\tablecaption{Disk Fractions\label{tab:diskfraction}}
\tablehead{
\colhead{$M_J$} &
\colhead{Chamaeleon I} &
\colhead{IC 348} &
\colhead{$\sigma$ Ori}}
\startdata
0-2 & 7/11 = $0.64^{+0.12}_{-0.15}$ & 7/19 = $0.37^{+0.12}_{-0.09}$ & 4/14 = $0.29^{+0.14}_{-0.09}$ \\
2-4 & 23/35 = $0.66^{+0.07}_{-0.09}$ & 19/43 = $0.44^{+0.08}_{-0.07}$ & 20/71 = $0.28^{+0.06}_{-0.05}$ \\
4-6 & 43/91 = $0.47^{+0.05}_{-0.05}$ & 57/139 = $0.41^{+0.04}_{-0.04}$ & 62/164 = $0.38^{+0.04}_{-0.04}$ \\
6-8 & 21/44 = $0.48^{+0.08}_{-0.07}$ & 34/71 = $0.48^{+0.06}_{-0.06}$ & 33/87 = $0.38^{+0.06}_{-0.05}$ \\
8-10 & 8/12 = $0.67^{+0.10}_{-0.15}$ & 11/19 = $0.58^{+0.10}_{-0.11}$ & 7/11 = $0.64^{+0.12}_{-0.15}$ \\
10-12 & 2/6 = $0.33^{+0.22}_{-0.12}$ & \nodata & 5/10 = $0.50^{+0.15}_{-0.14}$ \\
\enddata
\tablecomments{The statistical errors in the disk fractions are computed in the
manner described by \citet{bur03}.}
\end{deluxetable}

\clearpage

\end{document}